\documentclass[12pt,english,floatfix,superscriptaddress,aps,prd,preprint,showkeys,nofootinbib]{revtex4}
\usepackage{amsmath}
\usepackage{amssymb}
\usepackage{amsbsy}
\usepackage{amsfonts}
\usepackage{amsopn}
\usepackage{amstext}
\usepackage{graphicx}
\usepackage[english]{babel}
\usepackage{color}
\usepackage{slashed}
\usepackage{esint}
\usepackage[dvips]{epsfig}
\usepackage[dvips]{graphicx}
\usepackage{float}
\usepackage{units}
\usepackage{textcomp}
\usepackage{wasysym}
\usepackage{hyperref}
\usepackage{slashed}
\usepackage{subcaption}
\usepackage{dcolumn}
\usepackage{babel}
\usepackage{csquotes}
\usepackage{color}
\usepackage{slashed}
\usepackage{simplewick}

\begin{document}

\title{One-loop radiative corrections in bumblebee-Stueckelberg model}


\author{Fernando M. Belchior}
\email{belchior@fisica.ufc.br}
\affiliation{Universidade Federal do Cear\'a (UFC), Departamento de F\'isica,\\ Campus do Pici, Fortaleza - CE, C.P. 6030, 60455-760 - Brazil.}

\author{Roberto V. Maluf}
\email{r.v.maluf@fisica.ufc.br}
\affiliation{Universidade Federal do Cear\'a (UFC), Departamento de F\'isica,\\ Campus do Pici, Fortaleza - CE, C.P. 6030, 60455-760 - Brazil.}
\affiliation{Departamento de F\'{i}sica Te\'{o}rica and IFIC, Centro Mixto Universitat de Valencia - CSIC. Universitat
de Valencia, Burjassot-46100, Valencia, Spain.}

\begin{abstract}
This work aims to study the radiative corrections in a vector model with spontaneous Lorentz symmetry violation, known in the literature as the bumblebee model. We consider such a model with self-interaction quadratic smooth potential responsible for spontaneous Lorentz symmetry breaking. The spectrum of this model displays a transversal nonmassive mode, identified as Nambu-Goldstone, and a massive longitudinal mode. Besides the Lorentz symmetry, this model also exhibits gauge symmetry violation. To restore the gauge symmetry, we introduce the Stueckelberg field and calculate the two-point function by employing the principal-value (PV) prescription. The result is nontransversal, leading to a massive excited mode.
\end{abstract}

\maketitle

\section{INTRODUCTION}

Symmetries play an important role in the description of nature at both classical and quantum level. In this context, the Lorentz symmetry appears as an essential feature in the building of general relativity and quantum field theory. Although both theories are well succeeded in prediction of a wide range of effects, there are issues in open as dark matter, accelerated universe expansion and a consistent theory of quantum gravity. Hence these theories might be a low-energy limit of a fundamental theory located at Planck scale, where the effects of quantum gravity would be observed. In this sense, an important question arises: could the Lorentz symmetry be violated in a fundamental theory? This question is not treated as mere speculation since many theories like string theory \cite{Kostelecky:1988zi}, non commutative theories \cite{Carroll:2001ws} and Horava Lifshitz \cite{Horava:2009uw} lead to Lorentz symmetry violation (LSV) at Planck scale.

A pioneering study on the spontaneous LSV mechanism arising from certain field solutions in string theories with non-zero expected value was carried out by Kostelecky and Samuel in Ref. \cite{Kostelecky:1988zi}. Since then, there has been growing interest in understanding how LSV affects the usual physics. These studies have led to the development of an effective field theory known as the Standard Model Extension (SME) \cite{Colladay:1998fq, Colladay:1996iz}, which consists of Standard Model terms and all possible terms that violate Lorentz symmetry while preserving the renormalizability of the theory.

As is well known in the literature, there are several distinct ways of implementing LSV \cite{Kostelecky:2022idz, Russell:2015gwa}. In the context of SME, the simplest one is explicit LSV, where constant parameters are directly added to the Lagrangian of the theory, such as in Carroll-Field-Jackiw electrodynamics \cite{Carroll:1989vb}, as well as in other CPT-odd extensions \cite{Myers:2003fd,Ferreira:2019lpu} and CPT-even theories \cite{Casana:2013nfx, Casana:2012vu, Casana:2018rhg}. Another approach is spontaneous LSV, which was mainly motivated by string theory, where vector and tensor fields from an underlying theory can acquire non-zero vacuum expectation values, defining a privileged direction in spacetime \cite{Kostelecky:2003fs}. Both LSV mechanisms occur in the particle frame, while coordinate invariance remains intact.

Among the many models that consider spontaneous LSV, the simplest one involves a vector field with self-interaction known as the bumblebee field. This model was first introduced in Ref. \cite{Kostelecky:1988zi} and has been investigated in different contexts, such as black holes \cite{Maluf:2020kgf, Maluf:2022knd, DCarvalho:2021zpf, Kanzi:2021cbg, Li:2020dln, Casana:2017jkc}, wormholes \cite{Ovgun:2018xys, Oliveira:2018oha}, warped spacetime \cite{Lessa:2021npz, Santos:2012if}, the G\"{o}del universe \cite{Santos:2014nxm, Jesus:2020lsv}, cosmology \cite{Maluf:2021lwh}, and classical and quantum aspects \cite{Hernaski:2014jsa, Maluf:2013nva, Maluf:2014dpa, Escobar:2017fdi, Maluf:2015hda, Assuncao:2017tnz, Delhom:2022xfo, Delhom:2020gfv, Delhom:2019wcm}. Additionally, it is also possible to build a model with spontaneous LSV using an antisymmetric 2-tensor field known as the Kalb-Ramond field \cite{Altschul:2009ae, Maluf:2018jwc, Aashish:2019ykb, Aashish:2018aqn, Lessa:2019bgi, Lessa:2020imi, Maluf:2021eyu, Maluf:2021ywn, Assuncao:2019azw}.

Concerning the photon sector of the SME, the modification induced by the bumblebee field allows us to study contributions arising from extra modes resulting from the spontaneous symmetry-breaking mechanism in an underlying fundamental theory. Phenomenological consequences can manifest as generating additional terms or modifications in the electromagnetic field equations, potentially leading to observable effects at both theoretical and experimental levels. 

The vector and tensor bumblebee models incorporate a kinetic gauge-invariant term and a self-interacting potential that spontaneously breaks Lorentz and gauge symmetries. As a result, an LSV background is established, associated with the vacuum expectation values (VEVs) for the fields, defining privileged directions. The expansion of these fields around their VEVs reveals a massless Nambu-Goldstone mode and a massive longitudinal excitation, resulting from the LSV background but lacking physical propagating properties \cite{Altschul:2005mu, Bluhm:2004ep, Bluhm:2007bd, Carroll:2009em, Bluhm:2008yt}.

In this work, we focus on the bumblebee model and its extension that involves the coupling of the bumblebee field to the Stueckelberg field, as proposed in Ref. \cite{Hernaski:2014jsa}. Although the bumblebee model has been mainly explored in the gravitational sector, there are still issues to be worked on in flat spacetime, such as radiative corrections.  Therefore, the main goal of this work is to investigate the bumblebee-Stueckelberg model in the context of effective theories and explore quantum corrections beyond the tree-level order. This topic has recently been addressed in the context of explicit LSV \cite{Kostelecky:2001jc, Casana:2013nfx, Jackiw:1999yp, Nascimento:2007rb, Ferrari:2018tps, Mariz:2011ed, Altschul:2022isc, BaetaScarpelli:2021dhz}, bumblebee model in flat spacetime \cite{Maluf:2015hda}, and the metric-affine bumblebee model \cite{Delhom:2020gfv, Delhom:2022xfo}, which is a modified gravity theory where the metric and connection are treated independently.

Throughout this work, we stand out that the bumblebee model lacks gauge invariance, which can be restored by introducing the Stueckelberg field. We derive the propagators for the $\beta_\mu$ field and the Stueckelberg field and demonstrate that a suitable choice of vectors can avoid tree-level instabilities in the bumblebee-Stueckelberg model. Finally, we investigate radiative corrections for this model and find that the two-point function of the $\beta_\mu$ field is not transverse, indicating that quantum corrections excite the massive modes.

This work is outlined as follows. In Sec. (\ref{lm}), we briefly introduce our theoretical model by starting from bumblebee model and then we consider a coupling between the bumblebee field and Stueckelberg field. In Sec.(\ref{olc}), we carry out the study of radiative corrections at one-loop in the bumblebee-Stueckelberg model. Finally, in Sec. (\ref{c}), our conclusion as well as future perspectives are present.


\section{THE THEORETICAL MODEL}\label{tm}
Our theoretical model is a vector model in which the Lorentz symmetry is spontaneously broken by the presence of a smooth quadratic potential, leading the vector field to assume a non-zero vacuum expectation value (VEV) $\langle B_\mu\rangle = b_\mu$. The model is described by the following Lagrangian density \cite{Maluf:2015hda, Hernaski:2014jsa}:
\begin{equation}
\mathcal{L}=-\frac{1}{4}B_{\mu\nu} B^{\mu\nu}- \frac{\lambda}{4}\left(B_\mu B^\mu-b^2\right)^2,
\label{l1}
\end{equation}
where $B_{\mu\nu}=\partial_\mu B_\nu-\partial_\nu B_\mu$ is the field strength for the bumblebee field $B_\mu$, $\lambda$ is a dimensionless positive constant, and $b^{2}$ is a positive constant with squared mass dimension. The above Lagrangian contains a Maxwell-like term and a smooth quadratic potential, which is responsible for spontaneous LSV besides being gauge noninvariant. In addition to smooth potential form, such a potential can take other forms as Lagrange-multiplier \cite{Bluhm:2007bd} and nonpolynominal potential \cite{Altschul:2005mu}. 

In order to study the main properties and free propagation of bumblebee field at tree-level order, we can consider a small perturbation $\beta_\mu$ around its VEV $b_\mu$ 
\begin{equation}
B_\mu=b_\mu+\beta_\mu.\label{shift}
\end{equation}
The excitation $\beta_{\mu}$ gives rise to two distinct modes. Firstly, two Nambu-Goldstone (NG) modes are perpendicular to the vacuum expectation value of the bumblebee field, and secondly, one massive or longitudinal mode \cite{Bluhm:2004ep,Bluhm:2007bd}. In the linear regime, the two NG modes can be identified as the two polarization modes of the photon. On the other hand, the massive mode manifests as a tachyonic ghost excitation, resulting in an unbounded below Hamiltonian for the model \cite{Carroll:2009em}. Nevertheless, appropriately selecting the initial conditions for the field configurations makes it possible to identify ghost-free regions in phase space where the Hamiltonian remains positive \cite{Bluhm:2008yt}.

One way to analyze the physical spectrum of the model is by calculating the Feynman propagator and extracting the dispersion relations through its pole structure.
By substituting the shift (\ref{shift}) into the Lagrange density (\ref{l1}), we can write
\begin{equation}
\mathcal{L}=\mathcal{L}_{free}+\mathcal{L}_{int},
\end{equation}
where
\begin{equation}
\mathcal{L}_{free}=-\frac{1}{4}\beta_{\mu\nu} \beta^{\mu\nu}-\lambda b_\mu b_\mu \beta^\mu\beta^\nu,
\label{l2}
\end{equation}and
\begin{equation}
\mathcal{L}_{int}=-\lambda \beta_\mu \beta^\mu \beta_\nu b^\nu-\frac{\lambda}{4}\beta_\mu \beta^\mu \beta_\nu\beta^\nu.
\label{l3}
\end{equation}

The first term $\mathcal{L}_{free}$ represents the bilinear Lagrangian, and the second term $\mathcal{L}_{int}$ is the interaction term that includes trilinear and quadrilinear terms. The bilinear Lagrangian (\ref{l2}) leads to the following propagator for $\beta_\mu$ field:
\begin{equation}
\Delta^{\mu\nu}(k)=-\frac{i}{k^2}\left[\eta^{\mu\nu}-\frac{(k^\mu b^\nu + k^\nu b^\mu)}{(b\cdot k)}+\frac{(k^2+2\lambda b^2)}{2\lambda(b\cdot k)^2}k^\mu k^\nu\right],
\label{P1}
\end{equation}
which is similar to the propagator of the Maxwell electrodynamics in nonlinear gauge \cite{Escobar:2017fdi}. As we can see, the propagator (\ref{P1}) has a simple pole $k^2=0$, leading to non-massive excitation. This excitation is transversal and can be associated with a Nambu-Goldstone mode. Furthermore, the LSV leads to a double pole $(b\cdot k)=0$, which can be interpreted as a non-physical mode as it has a dispersion relation that leads to instabilities \cite{Maluf:2015hda}.

The study of radiative corrections of the bumblebee model was carried out in Ref. \cite{Maluf:2015hda}. Such a model is characterized by nontraversal self-energy $p_\mu\Pi^{\mu\nu}\neq0$, showing that the massive mode is excited by the interaction terms. The bilinear Lagrangian has a Proca-like term that violates the gauge symmetry besides the Lorentz symmetry. In order to restore the gauge symmetry, we introduce the auxiliary field through the following transformation 
\begin{equation}
\beta_\mu\rightarrow\beta_\mu - \frac{1}{\sqrt{\lambda}}\partial_\mu\phi,
\end{equation}
which is invariant under the following gauge transformation
\begin{align}
\beta_\mu\rightarrow\beta_\mu+\partial_\mu\theta, \\
\phi\rightarrow\phi+\sqrt{\lambda}\theta,
\end{align}
where $\theta$ is the gauge parameter. This procedure is commonly applied to massive vector fields, as is the example of the Maxwell-Proca model, where there is an incompatibility between renormalizability by power counting and negative norm states present in the spectrum of the theory. In Ref. \cite{Ferreira:2020wde}, the Stueckelberg procedure was studied in the context of the Carroll-Field-Jackiw model. Moreover, in Ref. \cite{Hernaski:2014jsa}, this procedure was used to solve instability problems and to make the canonical quantization to the free bumblebee field. In any case, the field $\phi$ works as an auxiliary field, i.e., it does not represent a physical degree of freedom.

Since the gauge symmetry was restored, a gauge fixing is needed and we adopt the following term \cite{Hernaski:2014jsa}
\begin{equation}
\mathcal{L}_{gf}=-\frac{1}{2\xi}\left(b^\mu b^\nu\partial_\mu\beta_\nu-2\xi\sqrt{\lambda}\phi\right)^2.
\end{equation}

The above gauge fixing term is convenient since it is cancels out the mixing between the $\beta_\mu$ and $\phi$. Besides, the parameter $\xi$ can be chosen conveniently. So, the effective bumblebee-Stueckelberg Lagrangian density can be written as 
\begin{equation}
\mathcal{L}= \mathcal{L}_{free}+\mathcal{L}_{3I}+\mathcal{L}_{4I},
\label{lm}
\end{equation}
where
\begin{equation} 
\mathcal{L}_{free}=-\frac{1}{4}\beta_{\mu\nu}\beta^{\mu\nu}-\lambda b_\mu b_\nu \beta^\mu\beta^\nu + \frac{1}{2\xi}b^\mu b^\nu b^\rho b^\sigma \beta_\mu \partial_\rho \partial_\sigma\beta_\nu - b_\mu b_\nu \partial^\mu\phi\partial^\nu\phi - 2\xi\lambda\phi^2,
\label{l4}
\end{equation}
\begin{equation}
\begin{split}
\mathcal{L}_{3I}&=-\lambda \beta_\mu \beta^\mu \beta_\nu b^\nu + \sqrt{\lambda}\beta_\mu \beta^\mu b^\nu \partial_\nu \phi + 2\sqrt{\lambda}\beta_\mu \beta_\nu b^\nu \partial^\mu\phi - 
\\&\beta_\mu b^\nu \partial^\mu  \partial_\nu \phi -\beta_\nu b^\nu \partial_\mu \phi\partial^\mu\phi  + \frac{1}{\sqrt{\lambda}} b^\nu \partial_\nu \phi \partial_\mu \phi\partial^\mu\phi,
\end{split}\label{L3i}
\end{equation}
\begin{equation}
\begin{split}
\mathcal{L}_{4I}&=-\frac{\lambda}{4}\beta_\mu \beta^\mu \beta_\nu\beta^\nu + \sqrt{\lambda}\beta_\mu \beta^\mu \beta_\nu\partial^\nu\phi - \frac{1}{2}\beta_\mu \beta^\mu \partial_\nu\phi \partial^\nu\phi-
\\& \beta_\mu\beta_\nu\partial^\mu\phi\partial^\nu\phi +\frac{1}{\sqrt{\lambda}}\beta_\mu\partial^\mu\phi\partial_\nu\phi\partial^\nu\phi -\frac{1}{4\lambda}\partial_\mu\phi\partial^\mu\phi \partial_\nu\phi\partial^\nu\phi,
\end{split}\label{L4i}
\end{equation}
where $\mathcal{L}_{free}$ is bilinear in the $\beta_{\mu}$ and $\phi$ fields, whereas $\mathcal{L}_{3I}$ and $\mathcal{L}_{4I}$ handle the trilinear and quadrilinear terms, respectively. As we can see from the above expressions, there are no mixing terms between the $\beta_\mu$ and $\phi$ fields, and the propagators can be obtained from the quadratic terms, as usual. Moreover, we can extract the vertices from the interaction terms in (\ref{L3i}) and (\ref{L4i}). In the next section, we will use these vertices to construct one-loop Feynman diagrams for the two-point function of the $\beta_\mu$ field. Therefore, the bilinear Lagrangian (\ref{l4}) yields the following propagator for the bumblebee field:
\begin{equation}
\Delta_{F}^{\mu\nu}(k)=\frac{i}{k^2}\left[-g^{\mu\nu}+\frac{k^\mu b^\nu+k^\nu b^\mu}{(b\cdot k)}-\frac{b^2 k^\mu k^\nu}{(b\cdot
 k)^2}\right]-i\xi\frac{k^\mu k^\nu}{(b\cdot k)^2 [(b\cdot k)^2+2\xi\lambda]}.
 \label{P2}
\end{equation}
The above propagator is the same as that of Maxwell electrodynamics in nonlinear gauge for $\xi=0$. It has a massless pole $k^2=0$ and a non-physical double pole $b\cdot k=0$, which also appears in the bumblebee model. There is still a pole $(b\cdot k)^2+2\xi\lambda=0$ which depends on gauge fixing. For Stueckelberg $\phi$ field, we get the following propagator
\begin{equation}
D(k)=\frac{-i}{(b\cdot k)^2+2\xi\lambda}.
\label{P3}
\end{equation}

It is easy to see that the Stueckelberg propagator also has a pole $(b\cdot k)^2+2\xi\lambda=0$. The non-physical modes that appear in propagators (\ref{P2}) and (\ref{P3}) lead to instabilities in the spectrum of the theory. However, we can handle these instabilities by choosing an appropriate set for $\beta_\mu$ modes. This choice is able to reduce the phase space of the bumblebee-Stueckelberg model to a phase space that is equivalent to the Maxwell electrodynamics in a nonlinear gauge. In fact, we choose the $\xi=0$ in the propagator (\ref{P2}), we obtain the propagator of the Maxwell model with a nonlinear gauge. The quantum equivalence between these theories was studied in Ref. \cite{Escobar:2017fdi}. For the free $\beta_\mu$ field, the solutions for can be written as follows \cite{Hernaski:2014jsa}:
\begin{align}
\beta_\mu^{(0)}(\vec{p})=\left(\frac{\vec{b}\cdot\vec{p}}{b_0}+\frac{\sqrt{-2\xi k}}{b_0},\vec{p}\right),\\
\beta_\mu^{(i)}(\vec{p})=\epsilon_\mu^{(i)}(\vec{p}), \hspace{1cm} i=1,2\\
\beta_\mu^{(3)}(\vec{p})=\left(\frac{\vec{b}\cdot\vec{p}}{b_0},\vec{p}\right),
\end{align}
where $\epsilon_\mu^{(i)}(\vec{p})$ are two space-like quadrivectors and are orthogonal to both $p_\mu^{(i)}(\vert\vec{p}\vert ,\vec{p})$ as for $b_\mu$. Using the properties of $\epsilon_\mu^{(i)}(\vec{p})$ we can derive the following expression
\begin{eqnarray}
\sum_{i=1}^2\epsilon_\mu^{(i)}\epsilon_\nu^{(i)}=\eta_{\mu\nu}+\frac{1}{b\cdot\overline{p}}(b_\mu\overline{p}_\nu+b_\nu\overline{p}_\mu)-\frac{1}{(b\cdot\overline{p})^2}\overline{p}_\mu\overline{p}_\nu.
\end{eqnarray}

It is worth mentioning that the above projector is the same that appears in the propagator for the field $\beta_\mu$ and, as we have already seen, this propagator is identical to the propagator of Maxwell electrodynamics in a non-linear gauge. In the next section, we will study the one-loop radiative corrections in the bumblebee-Stueckelberg model. Of course,

\section{ONE-LOOP CORRECTIONS}\label{olc}

In this section, we will calculate the divergent contributions of the one-loop radiative corrections for the present model. Our objective is to study the two-point Green function of the $\beta_{\mu}$ field and determine whether the introduction of the Stueckelberg field can render the theory renormalizable and eliminate non-physical modes. To achieve this goal, we will examine whether the two-point Green function of the field is transverse or exhibits one excited massive mode.

\begin{figure}[!h]
\begin{center}
\begin{tabular}{cc}
\includegraphics[scale=0.6]{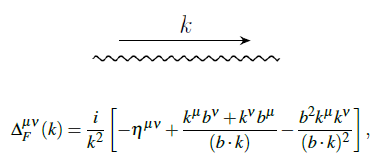}
\includegraphics[scale=0.6]{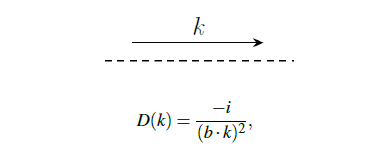}\\
\end{tabular}
\end{center}
\caption{Propagators for bumblebee-Stueckelberg model\label{fig1}}
\end{figure}

\begin{figure}[th!]
\begin{center}
\begin{tabular}{cc}
\includegraphics[scale=0.58]{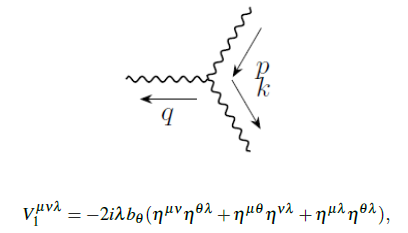}
\includegraphics[scale=0.58]{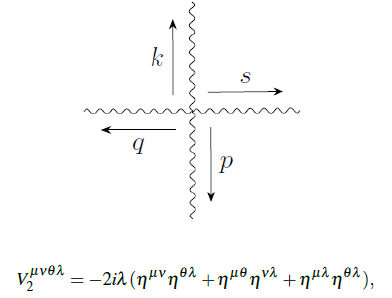}\\
\includegraphics[scale=0.58]{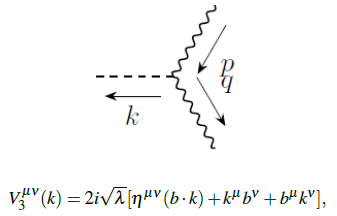}
\includegraphics[scale=0.58]{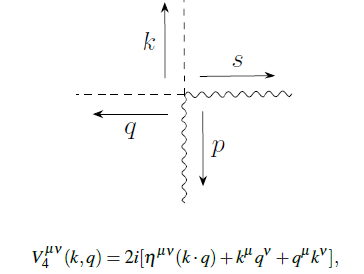}\\
\includegraphics[scale=0.58]{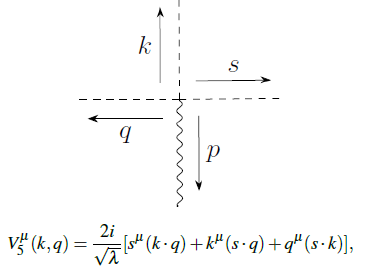}
\includegraphics[scale=0.58]{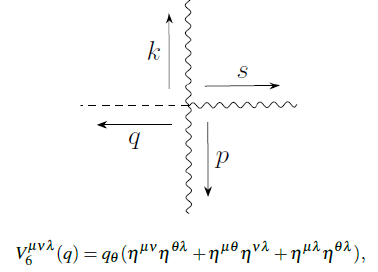}\\
\includegraphics[scale=0.58]{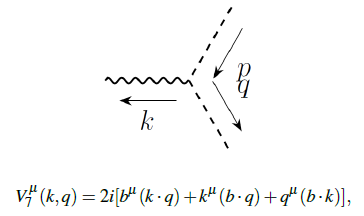}
\includegraphics[scale=0.58]{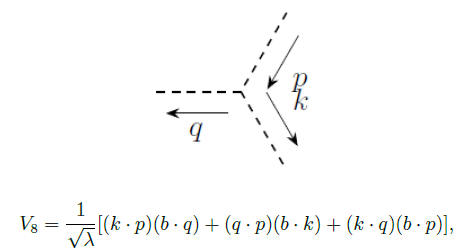}\\
\includegraphics[scale=0.58]{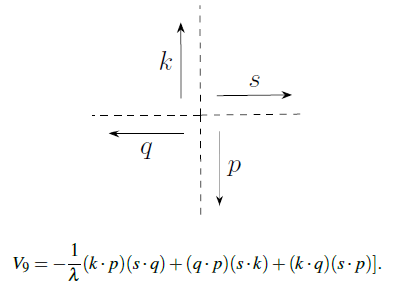}
\end{tabular}
\end{center}
\caption{Vertices for bumblebee-Stueckelberg model\label{fig2}}
\end{figure}

By adopting a perturbative approach, we can derive the Feynman rules from the classical Lagrangian defined in Eq. (\ref{lm}), which can be read off from Figs. (\ref{fig1}) and (\ref{fig2}). For our theoretical calculations, we restrict ourselves to the one-loop calculation to the self-energy function and fix the value of $\xi$ at the Landau gauge, namely, $\xi=0$. 

Once the Feynman rules have been determined, we can investigate radiative corrections. From Fig. \ref{Fig3}, there are five Feynman diagrams that contribute to the two-point $\Pi^{\mu\nu}$ Green function at one-loop order, allowing us to express $\Pi^{\mu\nu}$ as follows:

\begin{figure}[h!]
\begin{center}
\begin{tabular}{cc}
\includegraphics[scale=0.5]{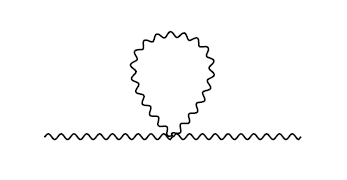}
\includegraphics[scale=0.5]{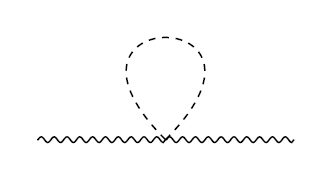}\\
(a) \hspace{4.1cm}  (b) \\
\includegraphics[scale=0.45]{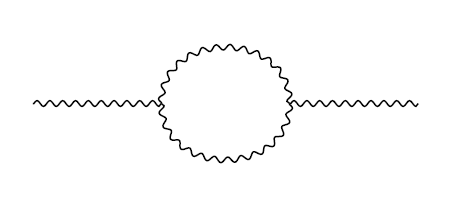}
\includegraphics[scale=0.45]{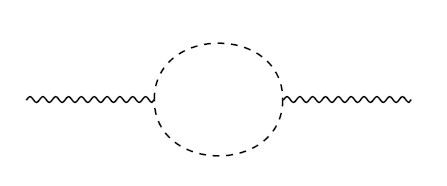}\\
(c) \hspace{4.8cm} (d)\\
\includegraphics[scale=0.45]{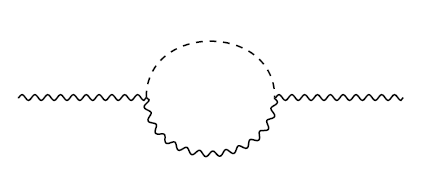}\\
(e)
\end{tabular}
\end{center}
\caption{(a) Tadpole massless, (b) Tadpole bumblebee-Stueckelberg, (c) One-loop bumblebee-pure, (d) One-loop Stueckelberg-pure e (e) One-loop bumblebee-Stueckelberg\label{Fig3}}
\end{figure}

\begin{equation}
\Pi^{\mu\nu}=\Pi^{\mu\nu}_a+\Pi^{\mu\nu}_b+\Pi^{\mu\nu}_c+\Pi^{\mu\nu}_d+\Pi^{\mu\nu}_e,
\end{equation}
where
\begin{equation}
\Pi^{\mu\nu}_a=\frac{1}{2}\int\frac{d^Dk}{(2\pi)^D}V_{2}^{\mu\nu\theta\lambda}\Delta_{\theta\lambda}(k),\label{pia}
\end{equation}

\begin{equation}
\Pi^{\mu\nu}_b=\frac{1}{2}\int\frac{d^Dk}{(2\pi)^D}V_{4}^{\mu\nu}D(k),\label{pib}
\end{equation}

\begin{equation}
\Pi_{c}^{\mu\nu}=\frac{1}{2}\int\frac{d^Dk}{(2\pi)^D}V_{7}^{\mu}(k,k+p)D(k)V_{7}^{\nu}(k,k+p)D(k+p),\label{pic}
\end{equation}

\begin{equation}
\Pi_{d}^{\mu\nu}=\frac{1}{2}\int\frac{d^Dk}{(2\pi)^D}V_{1}^{\mu\rho\lambda}(k)\Delta_{\rho\theta}(k)V_{1}^{\nu\sigma\theta}(k+p)\Delta_{\lambda\sigma}(k+p),\label{pid}
\end{equation}

\begin{equation}
\Pi_{e}^{\mu\nu}=\frac{1}{2}\int\frac{d^Dk}{(2\pi)^D}V_{3}^{\mu\rho}(k)D(k)V_{3}^{\nu\sigma}(k+p)\Delta_{\rho\sigma}(k+p).\label{pie}
\end{equation}

It is worth noting that, based on simple power-counting arguments, the one-loop integrals mentioned above may exhibit ultraviolet divergences up to fourth order. Moreover, due to the presence of non-physical pole $(b\cdot k)$ in the above integrals we need a consistent prescription to extract only the physical content of the theory. In the context of Yang-Mills theories there is a prescription that has been widely employed to deal with these pole the so-called principal-value (PV) prescription \cite{Leibbrandt:1983pj}. Such a prescription is defined as
\begin{equation}
\frac{1}{(b\cdot k)^\beta}=\frac{1}{2}\lim_{\mu\rightarrow 0} \left[\frac{1}{(b\cdot k+i\mu)^\beta}+\frac{1}{(b\cdot k-i\mu)^\beta}\right].
\end{equation}

The PV prescription is consistent with the unitarity and renormalization properties of Yang-Mills theories besides preserving The Slavnov-Taylor indentities at one-loop. The corresponding analytical expressions for one-loop integrals are shown in Appendix \ref{App}. We use the dimensional regularization in the PV prescription to evaluate the divergent contributions of momentum integrals. Adding up Eqs. (\ref{pia})-(\ref{pie}), and carrying out some algebraic manipulations, we arrive at
\begin{equation}
\Pi^{\mu\nu}=\frac{4\lambda}{3}\left(\frac{b\cdot p}{b^2}(b^\mu p^\nu + b^\nu p^\mu)+\frac{5[(b\cdot p)^2-2b^2p^2]}{b^4}b^\mu b^\nu\right)I_{div},
\label{TP}
\end{equation}
where $I_{div}=\frac{i}{8\pi^2\epsilon}$ and $\epsilon=4-D$. 

The two-point function in the bumblebee-Stueckelberg model, like in the bumblebee model, is not transverse, indicating the existence of a massive excited mode. This implies that the massive longitudinal mode remains a loop propagating mode even in a gauge-invariant theory, similar to the case in the absence of the Stueckelberg field \cite{Maluf:2015hda}. In addition, the one-loop correction does not give rise to any divergent term that is non-local or contains higher derivatives with respect to the external momentum. In future work, it would be interesting to investigate whether this non-transversality persists when considering the matter sector. To accomplish this, we intend to examine the two-point function by introducing an appropriate coupling between the bumblebee and matter fields (scalar and spinorial).

We can analyze the two-point function (\ref{TP}) in terms of propagating modes. For this purpose, we will define the longitudinal projector $P_{\mu\nu}^{\parallel}$ and the transverse projector $P_{\mu\nu}^{\perp}$, where
\begin{equation}
P_{\mu\nu}^{\parallel}=\frac{b_\mu b_\nu}{b^2},
\end{equation}

\begin{equation}
P_{\mu\nu}^{\perp}=\eta_{\mu\nu}-\frac{b_\mu b_\nu}{b^2}.
\end{equation}
Once the orthogonal projectors are defined, we can decompose the field $\beta_\mu$ into transversal mode $A_\mu$ and longitudinal mode $\beta \widehat{b}_\mu$

\begin{equation}
\beta_\mu=A_\mu+\beta\widehat{b}_\mu,
\label{E}
\end{equation} 
where
\begin{eqnarray}
A_\mu=P_{\mu\nu}^{\perp}\beta^\nu,\\
\beta \widehat{b}_\mu=P_{\mu\nu}^{\parallel}\beta^\nu.
\end{eqnarray}
with $A_\mu b^\mu=0$, $\widehat{b}_\mu=\frac{b_\mu}{b}$ and $\widehat{b}_\mu\widehat{b}^\mu=1$. As we said before, the transverse mode $A_{\mu}$ represents the NG excitations and is associated with the photon mode. Besides, the longitudinal mode $\beta$ is associated with massive excitation. Therefore, by employing the decomposition (\ref{E}), the free Lagrangian (\ref{l4}) becomes
\begin{equation}
\begin{split}
\mathcal{L}_{free}=&-\frac{1}{4}F_{\mu\nu}F^{\mu\nu}-\frac{1}{2}F^{\mu\nu}\partial_{[\mu}\beta \widehat{b}_{\nu]}-\frac{1}{2}\partial_\mu\beta\partial^\mu\beta +\frac{1}{2}\partial_\mu\beta\partial_\nu\beta\widehat{b}^\mu \widehat{b}^\nu\\&  +\frac{1}{\xi}b^2\beta(b\cdot\beta)^2\beta - \lambda b^2\beta^2 - b_\mu b_\nu \partial^\mu\phi\partial^\nu\phi - 2\xi\lambda\phi^2.
\end{split}
\end{equation}
Therefore, the contribution of self-energy (\ref{TP}) to effective action is given by

\begin{equation}
\mathcal{L}_{div}=\frac{\lambda}{3\pi^2\epsilon}\left[-\frac{1}{2}F^{\mu\nu}\partial_{[\mu}\beta \widehat{b}_{\nu]}+\frac{5}{2}\partial_\mu\beta\partial^\mu\beta -\frac{8}{2}\partial_\mu\beta\partial_\nu\beta\widehat{b}^\mu \widehat{b}^\nu\right],
\end{equation}
showing that we can obtain counter-terms from the free Lagrangian to renormalize the divergent term in Eq. (\ref{TP}).

\section{CONCLUSION}\label{c}

We have studied one-loop radiative corrections by considering the bumblebee field coupled to the Stueckelberg field. To begin with, we reviewed the bumblebee model with self-interaction due to a smooth potential that violated both Lorentz symmetry and gauge symmetry. From the propagator for $\beta_\mu$, we showed that, as is characteristic of models with global symmetry violation, a non-massive transverse mode was present, which could be identified as a Nambu-Goldstone mode. Additionally, a massive longitudinal mode appeared due to LSB, which had a non-physical behavior. Moreover, we observed that the self-energy of the $\beta_\mu$ field was not transverse, indicating that the self-interaction terms of the bumblebee field excited the massive mode.

In order to restore gauge symmetry, we introduced the Stueckelberg field and calculated the free-propagator for both the $\beta_\mu$ field and the Stueckelberg field. From there, we extracted Feynman rules and computed the two-point field function in the bumblebee-Stueckelberg model, showing that the self-energy was also not transverse. It is worth mentioning that when the $\beta_\mu$ field was coupled to the Stueckelberg field, its propagator became identical to the propagator of electrodynamics in the nonlinear gauge. In contrast, in the bumblebee model without the presence of the Stueckelberg field, the propagator depended on the coupling constant and had a shape similar to the propagator of Maxwell theory in the nonlinear gauge.

In the phenomenological context, it is important to point out experimental measurements, such as precise polarization measurements of photons, conducted by experiments like the Fermi Gamma-ray Space Telescope (Fermi-LAT) \cite{Fermi-LAT:2009ihh}, have placed constraints on photon polarization dispersion relations, providing insights into potential modifications in the photon sector of QED. Additionally, observations of high-energy cosmic rays, particularly ultra-high-energy cosmic rays (UHECRs), and astrophysical phenomena such as gamma-ray bursts (GRBs) and blazars \cite{Murase:2011cy, Piran:2004ba}, have been extensively analyzed to explore possible Lorentz symmetry violations in the photon sector. Cavity-based tests utilizing optical cavities have also been utilized to investigate modifications to properties of photons, such as the speed of light, providing further constraints on potential Lorentz-violating effects in the photon sector of the SME \cite{Muller:2002uk, Muller:2004zp, Kostelecky:2001mb}.

As a future perspective, our intention is to study radiative corrections in the bumblebee-Stueckelberg model by including coupling with matter and discussing a suitable renormalization scheme for this theory. We also aim to investigate the quantum-level equivalence between the bumblebee-Stueckelberg model and Maxwell electrodynamics in the nonlinear gauge. Additionally, we plan to explore the Kalb-Ramond model with LSV, which is a natural generalization of the bumblebee vector model.

\section*{Acknowledgments}
\hspace{0.5cm} The authors thank the Conselho Nacional de Desenvolvimento Cient\'{i}fico e Tecnol\'{o}gico (CNPq), Grants no. 200879/2022-7 (RVM) for financial support. R. V. Maluf acknowledges the Departament de F\'{i}sica Te\`{o}rica de la Universitat de Val\`{e}ncia for the kind hospitality.

\appendix

\section{AMPLITUDES\label{App}}

In this Appendix, we will show the analytic expressions for the one-loop self-energy. Such expressions are written below

\begin{equation}
\begin{split}
\Pi^{\mu\nu}_a &=\frac{1}{2}\int\frac{d^Dk}{(2\pi)^D}V_{2}^{\mu\nu\theta\lambda}\Delta_{\theta\lambda}(k) \\&=-i\lambda\int\frac{d^Dk}{(2\pi)^D}[\eta^{\mu\nu}\Delta^{\rho}_\rho(k) + 2\Delta^{\mu\nu}(k)]
\\&=\lambda\int \frac{d^D k}{(2\pi)^D}\left\lbrace\frac{-\eta^{\mu\nu}D}{k^2}-\left(1+\frac{2}{D}\right)\frac{\eta^{\mu\nu}b^2}{(b\cdot k)}+\frac{2(k^\mu b^\nu+k^\nu b^\mu)}{k^2(b\cdot k)}\right\rbrace
\end{split},
\end{equation}

\begin{equation}
\begin{split}
\Pi^{\mu\nu}_b &=\frac{1}{2}\int\frac{d^Dk}{(2\pi)^D}V_{4}^{\mu\nu}D(k)\\
&=\int \frac{d^D k}{(2\pi)^D}\frac{(\eta^{\mu\nu}k^2+2k^\mu k^\nu)}{(b\cdot k)^2}\\&=\left(1+\frac{2}{D}\right)\eta^{\mu\nu}\int\frac{d^D k}{(2\pi)^D}\frac{k^2}{(b\cdot k)^2}
\end{split},
\end{equation}

\begin{equation}
\begin{split}
\Pi_{c}^{\mu\nu} &=\frac{1}{2}\int\frac{d^Dk}{(2\pi)^D}V_{7}^{\mu}(k,k+p)D(k)V_{7}^{\nu}(k,k+p)D(k+p)\\
&=\int \frac{d^D k}{(2\pi)^D}[k^\mu k^\nu\Pi_{kk} + p^\mu p^\nu\Pi_{pp} + b^\mu b^\nu\Pi_{bb} + b^{(\mu} k^{\nu)}\Pi_{bk}\\& + k^{(\mu} p^{\nu)}\Pi_{kp} + b^{(\mu} p^{\nu)}\Pi_{bp}]
\end{split},
\end{equation}
where

\begin{equation}
\Pi_{pp}=\frac{2}{[b\cdot(k+p)]^2},
\end{equation}

\begin{equation}
\Pi_{bp}=\frac{2[k\cdot(k+p)]}{(b\cdot k)[b\cdot(k+p)]^2},
\end{equation}

\begin{equation}
\Pi_{bb}=\frac{2[k\cdot(k+p)]^2}{(b\cdot k)^2[b\cdot(k+p)]^2},
\end{equation}

\begin{equation}
\Pi_{kp}=\frac{2}{[b\cdot(k+p)]^2}+\frac{2}{(b\cdot k)[b\cdot(k+p)]},
\end{equation}

\begin{equation}
\Pi_{bk}=\frac{2(k+p)^2}{(b\cdot k)[b\cdot(k+p)]^2}+\frac{2[k\cdot(k+p)]}{(b\cdot k)[b\cdot(k+p)]^2},
\end{equation}

\begin{equation}
\Pi_{kk}=\frac{2}{(b\cdot k)^2}+\frac{4}{(b\cdot k)[b\cdot(k+p)]}+\frac{2}{[b\cdot(k+p)]^2}.
\end{equation}

\begin{equation}
\begin{split}
\Pi_{d}^{\mu\nu} &=\frac{1}{2}\int\frac{d^Dk}{(2\pi)^D}V_{1}^{\mu\rho\lambda}(k)\Delta_{\rho\theta}(k)V_{1}^{\nu\sigma\theta}(k+p)\Delta_{\lambda\sigma}(k+p)\\
&=\int \frac{d^D k}{(2\pi)^D} b^\mu b^\nu\Pi_{bb},
\end{split}
\end{equation}

where

\begin{equation}
\begin{split}
\Pi_{bb}&=\frac{2\lambda^2 (D-2)}{k^2(k+p)^2}+\frac{2\lambda^2 b^2}{k^2[b\cdot(k+p)]^2}+\frac{2\lambda^2 b^2}{(b\cdot k)^2(k+p)^2}\\&-\frac{4\lambda ^2 b^2[k\cdot(k+p)]}{k^2(b\cdot k)(k+p)^2[b\cdot(k+p)]}+\frac{2\lambda ^2 b^4[k\cdot(k+p)]^2}{k^2(b\cdot k)^2(k+p)^2[b\cdot(k+p)]^2}
\end{split}.
\end{equation}

\begin{equation}
\begin{split}
\Pi_{e}^{\mu\nu} &=\frac{1}{2}\int\frac{d^Dk}{(2\pi)^D}V_{3}^{\mu\rho}(k)D(k)V_{3}^{\nu\sigma}(k+p)\Delta_{\rho\sigma}(k+p)\\
&=\int \frac{d^D k}{(2\pi)^D}[\eta^{\mu\nu}\Pi_{\eta}+k^\mu k^\nu\Pi_{kk} + p^\mu p^\nu\Pi_{pp} + b^\mu b^\nu\Pi_{bb} + b^{(\mu} k^{\nu)}\Pi_{bk}\\&+ k^{(\mu} p^{\nu)}\Pi_{kp} + b^{(\mu} p^{\nu)}\Pi_{bp}]
\end{split},
\end{equation}

where

\begin{equation}
\Pi_{\eta}=\frac{2\lambda}{(k+p)^2},
\end{equation}

\begin{equation}
\Pi_{kk}=\frac{2\lambda b^2}{(k+p)^2[b\cdot(k+p)]^2},
\end{equation}

\begin{equation}
\Pi_{pp}=\frac{2\lambda b^2}{(k+p)^2[b\cdot(k+p)]^2},
\end{equation}

\begin{equation}
\Pi_{bb}=\frac{2\lambda k^2}{(b\cdot k)^2(k+p)^2}-\frac{8\lambda[k\cdot(k+p)]}{(b\cdot k)(k+p)^2[k\cdot(k+p)]}+\frac{2\lambda b^2[k\cdot(k+p)]}{(b\cdot k)^2(k+p)^2[b\cdot(k+p)]^2},
\end{equation}

\begin{equation}
\Pi_{kp}=\frac{2\lambda b^2[k\cdot(k+p)]}{(k+p)^2[b\cdot(k+p)]^2},
\end{equation}

\begin{equation}
\Pi_{kb}=-\frac{4\lambda}{(k+p)^2[b\cdot(k+p)]}+\frac{2\lambda}{(b\cdot k)^2[b\cdot(k+p)]}+\frac{2\lambda b^2[k\cdot(k+p)]}{(b\cdot k)(k+p)^2[b\cdot(k+p)]^2},
\end{equation}

\begin{equation}
\Pi_{bp}=-\frac{4\lambda}{(k+p)^2[b\cdot(k+p)]}+\frac{2\lambda b^2[k\cdot(k+p)]}{(b\cdot k)(k+p)^2[b\cdot(k+p)]^2}.
\end{equation}

\end{document}